\documentclass{aa}  
\usepackage{graphicx}
\usepackage{txfonts}
\usepackage{hyperref}
\usepackage{rotating}
\newboolean{acceptcomments}
\setboolean{acceptcomments}{true}  
\newcommand{\comment}[2]{%
  \ifthenelse{\boolean{acceptcomments}}%
    {\textcolor{#1}{#2}}  
    {}  
}

\begin{document} 
   \title{Determination of [Fe/H] in Fundamental mode Classical Cepheids from  Gaia DR3 light curve Fourier parameters}


   \author{R. Molinaro
          \inst{1}
          V. Ripepi\inst{1}
          M. Marconi\inst{1}
          M. Gatto\inst{1}
          E. Trentin\inst{1}
          M. Deka\inst{1}
          G. De Somma\inst{1, 4}
          I. Musella\inst{1}
          T. Sicignano\inst{2,3,1,4}
          F. Ficara\inst{1,5}
          S. Leccia\inst{1}
          \and
          E. Luongo\inst{1,5}
          }

   \institute{INAF-OACN Osservatorio Astronomico di Capodimonte, Salita Moiariello 16, 80131, Napoli (ITALY)\\
              \email{roberto.molinaro@inaf.it} \and European Southern Observatory, Karl-Schwarzschild-Strasse 2, 85748 Garching bei München, Germany 
   \and Scuola Superiore Meridionale, Largo San Marcellino10 I-80138 Napoli, Italy Napoli, Italy \and Istituto Nazionale di Fisica Nucleare, Sez. di Napoli, Monte S. Angelo, Via Cinthia Edificio 6 I-80126 Napoli, Italy
   \and 
   Università di Salerno, Dipartimento di Fisica “E.R. Caianiello”, Via Giovanni Paolo II 132, 84084 Fisciano (SA), Italy
             }

   \date{Received September 15, 1996; accepted March 16, 1997}

 
  \abstract
   {Estimating the iron abundance of Classical Cepheids (CCs) holds dual significance for stellar population studies. First, understanding the role of metallicity is crucial for defining the Period-Luminosity (PL) relation, a fundamental step in the cosmic distance scale. Secondly, as CCs are reliable tracers of young stellar populations, knowledge of their individual metallicities provides a robust  estimate of the chemical content of the stellar systems or regions where they are observed. However, estimating their metallicity remains challenging, as it relies primarily on time-consuming spectroscopic observations.}
   {The primary objective of this work is to determine if Fourier parameters, extracted from the multi-band photometry provided by the $Gaia$ mission, can serve as proxies for the star's iron abundance ([Fe/H]), to establish a reliable correlation between these pulsational properties and chemical composition.}
   {The reference sample comprises 398 CCs from Gaia DR3 for which high-resolution spectroscopic metallicities are available in the literature, and covering the typical pulsation period range of Classical Cepheids. Focusing exclusively on fundamental mode pulsators, each light curve was modelled with a truncated Fourier series to derive characteristic parameters, such as amplitude ratios and phase differences. A Random Forest algorithm was applied to this reference set as a feature selection tool to identify the most influential predictors, which were subsequently used to derive an empirical relation for estimating [Fe/H] from photometric properties.}
   {We inferred an empirical relation that combines characteristic Fourier parameters extracted from the multi-band Gaia light curves to enable the estimation of the [Fe/H] of fundamental mode Classical Cepheids. As a consistency check, we applied this relation to an independent target sample of Cepheids in the Milky Way (MW), the Large Magellanic Cloud (LMC), and the Small Magellanic Cloud (SMC). The resulting metallicity distributions exhibit three clearly separated peaks, centred near the currently accepted [Fe/H] values of these stellar systems, supporting the predictive capability and external validity of the derived relation.}
   {}

    \authorrunning{R. Molinaro et al.}
 \titlerunning{Photometric [Fe/H] in Classical Fundamental mode Cepheids}
      \maketitle

\section{Introduction}
Classical Cepheids (CCs) stand as fundamental pillars in astrophysics, serving as crucial primary distance indicators, thanks to the tight correlation between their pulsation period and intrinsic luminosity, known as the period-luminosity (PL) relation, or Leavitt Law \citep{lea12}. Accurate calibration of this relation is essential for defining the cosmic distance ladder, in particular to address the existing tension (dubbed Hubble tension) between the local value of the Hubble constant \citep[$\rm H_0$,][]{rie22,rie24} and the value predicted by the $\Lambda$CDM model combined with the Cosmic Microwave Background data \citep[][and references therein]{planck2020}. Addressing this tension requires the study  of potential systematic uncertainties in the PL relation, particularly the dependence of the relation on the star metallicity \citep[see e.g.][]{rip21, rip22, bre22,bre24,des22,tre24a,bha24,bre25,rip25}

Beyond their role in measuring distances, CCs are robust tracers of young stellar populations, and crucially, the metallicity of a CC is representative of the chemical content of the stellar system or the region of the galaxy where it was formed. Therefore, accurate knowledge of the metallicity of large CC samples, both within the Milky Way (MW) and in extragalactic systems such as the Magellanic Clouds, is fundamental for studying galactic chemical evolution and constraining theoretical models of metal distribution over time \citep{lem13, gri18, cat24, lon25}. 

The conventional and most reliable method for determining the metallicity of CCs is high-resolution spectroscopy. However, this method is inherently time-consuming, limiting its application primarily to relatively bright stars and relatively small samples. In the era of massive photometric surveys, such as the Gaia mission or the forthcoming Rubin-LSST survey, the development of reliable alternative and relatively faster methods for metallicity estimation is demanding. Such methods would allow the metallicity content of an exceptionally large sample of variables, including CCs in external galaxies, to be determined rapidly, thereby greatly speeding up all studies limited by knowledge of chemical abundances, such as pulsation analyses and models of stellar chemical evolution.

Observationally, the iron abundance ([Fe/H]) serves as a robust proxy for the overall stellar metallicity. Several studies in the literature have examined a potential connection between [Fe/H] and light-curve morphology. For RR Lyrae stars, pioneering works initially relied on limited samples and single passband photometry. For instance, \citet{sim88} established an empirical correlation between [Fe/H] and the second order Fourier phase difference ($\phi_{21}$) using Johnson $V$-band light curves. Around the same period, alternative approaches explored relationships between metallicity and radial velocity curves, though these studies were inherently severely limited by small sample sizes \citep{pon01}

Subsequent studies refined these early photometric calibrations. \citet{kov95} derived  polynomial relations combining the pulsation period with either two or four $V$-band Fourier parameters to predict [Fe/H], while \citet{jur96} formulated a  linear relation linking [Fe/H] directly to the period and $\phi_{31}$ in the $V$ band. More recently, the field shifted towards near-infrared wavelengths and advanced statistical methods. \citet{dek21} applied machine-learning regressions to OGLE $I$ band Fourier parameters, delivering a relation for RRab stars based on P, $\phi_{31}$, and $A_2$, and a separate equation for RRc stars involving P, $A_1$, $A_2$, and $\phi_{31}$. Along the same lines, \citet{mul21} extensively investigated correlations between $I$-band light-curve descriptors and spectroscopic abundances.

With the advent of space-based astrometry, the first comprehensive attempt to connect [Fe/H] with Fourier parameters utilizing Gaia data was carried out by \citet{ior21} within the context of halo structures. They obtained two relations for RRab and RRc based on the pulsational period and the G band Fourier phase differences $\phi_{31}$. Finally, a very recent work by \citet{mur25} successfully extended photometric metallicity calibrations to the Gaia DR3 RR Lyrae catalog. Specifically they delivered a formulation based on the period and the G-band $\phi_{31}$ parameter for the fundamental-mode RRab variables, and a multivariate relation incorporating the period, the G band $\phi_{31}$, and the second order G band amplitude $\rm A_2$ for the first overtone RRc pulsators.

For Classical Cepheids, early photometric metallicity calibrations similarly relied on single band formulations in the optical domain. \citet{zso95} first attempted to calibrate [Fe/H] using standard UBV photometry, deriving an empirical relation that combined the parameters $\rm A_1, \phi_{21}, R_{31}, \phi_{31}, R_{51}$, although the calibration was strictly constrained by a very limited sample of stars. Subsequently, \citet{kla13} expanded this approach by analyzing light curve parameters across multiple individual filters, delivering single band relations based on the amplitude ratios $\rm R_{21}$ and $\rm R_{31}$. However, these formulations were characterized by tight applicability boundaries in terms of pulsation period. Using the near-infrared regime, \citet{sko16} implemented a multi-band framework using space-based observations from Spitzer and the James Webb Space Telescope. They calibrated [Fe/H] as a function of the color difference derived by subtracting a baseline LMC period-color relation from the observed color of the target, though their method remained valid only within a restricted period window between 6 and 90 days. More recently, optical calibrations were revisited by \citet{szi18}, who established a polynomial relation in the Johnson $V$ band linking the iron abundance to the parameter subset $\rm A_1, R_{21}, \phi_{21}, R_{31},$ and $\phi_{31}$. In a very recent work, \citet{hoc23}, provided linear photometric relations for fundamental-mode Classical Cepheids using both $V$ and $I$ band from the OGLE survey. They separated their analysis into distinct short-period ($2.5 < P < 6.3$ days) and long-period ($12 < P < 40$ days) regimes. For the short-period stars, they demonstrated that the best fit is achieved by a combination of the  amplitudes $\rm A_1$ and $\rm A_2$, achieving an intrinsic scatter of 0.12 dex. For the long-period domain, their best fit metallicity relation combines the amplitude $\rm A_1$, the phase difference $\phi_{21}$, and the fourth-order amplitude ratio $\rm R_{41}$, yielding a dispersion of 0.25 dex.

Nevertheless, despite all these efforts, a reliable empirical framework specifically tailored to the homogeneous, multi band photometry of the Gaia mission is still completely missing for Classical Cepheids. Given this context, the primary objective of the present work is to bridge this gap by developing a robust photometric relation calibrated directly onto the Gaia DR3 dataset.

The European Space Agency's $Gaia$ mission \citep{gaia2016}, with its third Data Release (DR3)\footnote{The Gaia data used in this work are publicly available from the ESA Gaia Archive at \url{https://gea.esac.esa.int/archive/}.}, has provided the scientific community with photometric time series in the $\rm G$, $\rm G_{BP}$, and $\rm G_{RP}$ bands for 15006 variable sources of the Cepheid type, of which 7712 are classified as fundamental mode CCs \citep{cle23,rip23}. Furthermore, precise spectroscopic estimates of the [Fe/H] abundance are available for a subset of these sources, derived from high-resolution spectroscopy measurements \citep{rom08, das22, kov22,  rom22, rip21, rip22, tre23, tre24b, rip25}. Building upon this unique dataset, our primary goal in this work is to establish a robust empirical relation linking the [Fe/H] values to the light-curve parameters across the different Gaia photometric bands. To this aim, we adopt a supervised machine-learning approach based on a Random Forest (RF) algorithm as a feature-selection tool to identify the photometric parameters that are most strongly correlated with [Fe/H] within a calibrating reference sample for which high-resolution spectroscopic abundances are available. The subset of parameters identified in this way is then used to derive an empirical linear relation for estimating the iron abundance. To assess the reliability and external validity of our results, we subsequently apply this calibrated relation to an independent target sample of CCs from the Gaia DR3 catalog, covering different galactic environments.

The paper is structured as follows: Section~\ref{sec:data} describes the reference spectroscopic sample used for calibration; the Fourier modeling of the light curves, the RF analysis used to identify the most influential predictors, and the subsequent linear fitting procedure are described in Section~\ref{sec-feh-fit}; the application of the derived relation to the independent target samples of the MW, LMC, and SMC, along with the discussion of the resulting metallicity distributions, is presented in Section~\ref{sec-application}; finally, our conclusions are summarized in Section~\ref{sec-conclusions}.

\section{The data}\label{sec:data}

The sample of fundamental mode CCs used in this work is based on data from the literature. Targets were selected to ensure: i) secure classification, with precise periods and photometry in the Gaia bands \citep{rip23}; ii) reliable and homogeneous [Fe/H] abundances from high-resolution spectroscopy; and iii) the widest possible range in [Fe/H]. We restrict the analysis to fundamental-mode pulsators, as first-overtone CCs generally show more regular and lower-amplitude light curves, resulting in a reduced sensitivity of light-curve shape parameters to metallicity.

As a reference, we adopted the work by \citet{tre24b}, which includes 182 Galactic F-mode CCs with homogeneous metallicities spanning [Fe/H] from approximately +0.3 to $-$1.0 dex. To complement this sample, we added 41 and 207 Galactic CCs with [Fe/H] abundances published by \citet{kov22} and \citet{das22}, respectively. Both studies share a significant number of objects with \citet{tre24b}. Since we adopt \citet{tre24b} as the reference scale, we used its metallicity values in cases of overlap. The common objects were, however, instrumental in homogenizing the \citet{kov22} and \citet{das22} abundances onto the \citet{tre24b} scale, as described in Appendix ~\ref{sec:metallicity_fit}.

Because the additional objects from \citet{kov22} and \citet{das22} span a relatively limited metallicity range, the combined Galactic sample was skewed toward solar metallicities. To mitigate this potential bias, we further extended the sample by including CCs from the LMC and SMC, with typical metallicities of [Fe/H] $\sim -0.4$ and $\sim -0.75$ dex, respectively. Specifically, we added 88 CCs in the LMC \citep[][]{rom22} and 14 CCs in the SMC \citep{rom08}.
We note that combining the Galactic sources from Trentin et al sample with the LMC and SMC datasets from \citet{rom08, rom22}\footnote{We are fully aware that \citet{rom22} acknowledged a potential selection bias in the \citet{rom08} sample. However, once those spectroscopic values are formally revised, they do not change significantly. This updated, yet unpublished, result is explicitly discussed and cited by \citet{bre24}.} does not introduce significant systematic scale mismatches. First, both the Milky Way and Magellanic Cloud spectra were predominantly secured using the same instrumental setup (VLT/UVES) and analyzed using the same atomic line lists and reference solar abundances. Second, while the \citet{rom22} pipeline employs a slightly modified approach for atmospheric parameter determination in lower-metallicity regimes, this framework was fully incorporated and verified by \citet{tre24a}. Their cross-calibration demonstrated excellent consistency and the absence of detectable zero-point offsets, ensuring that the combined reference sample provides a homogeneous and reliable spectroscopic scale spanning from $\rm [Fe/H] \sim -1.0$ to $+0.5$ dex.

The initial sample consists of 532 fundamental mode CCs.



\subsection{Fourier analysis}\label{sec:fourier-analysis}
The light curves in the G, $\rm G_{BP}$, and $\rm G_{RP}$ bands were modelled using a truncated Fourier series. This approach allows us to represent the periodic magnitude variation m(t) as a superposition of a fundamental frequency and its harmonics, as shown in the following equation:
\begin{equation}
m(t) = m_0 + \sum_{i=1}^{N} A_i \cos\left[ \frac{2\pi k}{P} (t - T_0) + \phi_i \right],
\end{equation}\label{eq:fourier_series}
where N is the number of harmonics, while P and $\rm T_0$ are the period and the epoch at maximum of light from $Gaia$ DR3, respectively.
The Fourier decomposition coefficients derived from this fit ($A_i$ and $\phi_i$) are subsequently used to define a set of characteristic Fourier parameters that describe the light curve shape. These parameters are crucial for studying relations between the light curve morphology and intrinsic physical properties, such as [Fe/H], which is the focus of the present work.

Specifically, we computed the following relative parameters, based on the amplitudes ($\rm A_i^X$) and phases ($\rm \phi_i^X$) of the fitted Fourier series in generic band X \citep{sim81}:
\begin{equation}\label{eq-fourier-params}
    \rm R^X_{i1} = \frac{A^X_i}{A^X_1} ;\newline
    \rm \phi^X_{i1} = \phi^X_i - i\cdot \phi^X_1.
\end{equation}
We focused our attention to the cases i=2,3, i.e. we considered $R_{21}^X$, $R_{31}^X$, $\phi_{21}^X$, $\phi_{31}^X$ for X = G, $\rm G_{BP}$, $\rm G_{RP}$.
This selection stems from our attempt to find an optimal compromise between maximizing the exploitation of light-curve morphology and ensuring the applicability of the relation to the largest possible number of sources. It is clear that increasing the truncation order of our relation to harmonics higher than the third order would have further restricted its applicability only to sources with exceptionally well-sampled photometric curves. Conversely, adopting a lower harmonic order would have meant failing to fully exploit the rich morphology of the light curves. Therefore, our choice represents a well balanced compromise.

In addition to the Fourier parameters defined above, we have also considered other characteristic light curve properties, such as the peak-to-peak amplitude and the pulsational period. The total number of considered parameters is equal to $N_{par}=25$.

From the initial sample, we further excluded 134 sources whose Fourier decomposition involved fewer than three harmonics ($n_{\rm harm} < 3$), thus not allowing a reliable computation of the Fourier parameters defined above.
The remaining sample contains 398 CCs and will be dubbed as the reference sample in the subsequent analysis. For the sake of reproducibility, the full list of the calibration stars along with their Gaia DR3 source identifiers, pulsation periods, reference spectroscopic metallicities, and key light curve quality parameters is provided in Table \ref{tab:ref_sample_param}.

\begin{figure}
   \centering
  \includegraphics[trim=0cm 1.0cm 0.0cm 2cm, clip, width=1.0\linewidth]{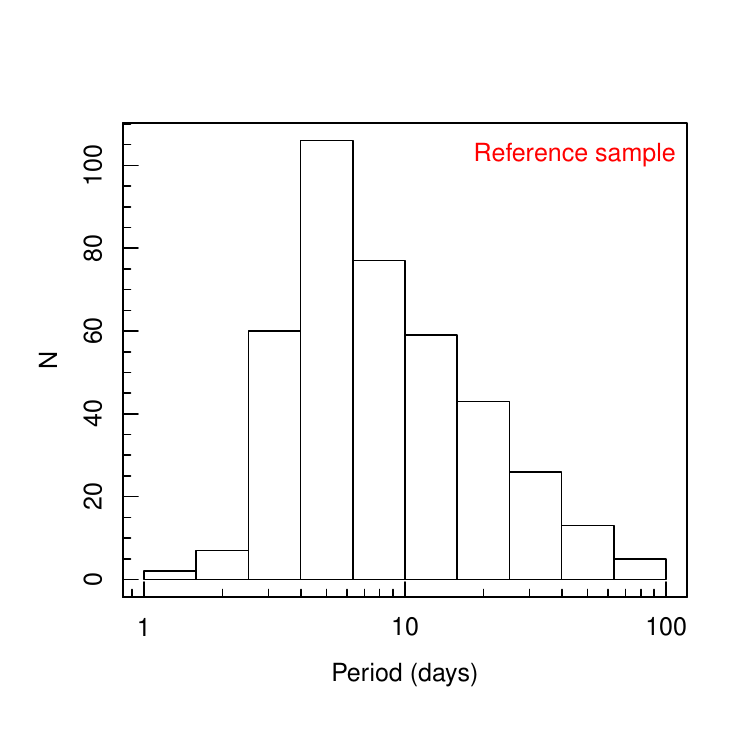}
   \caption{Period distribution of the 398 CCs constituting the reference sample. The periods range from approximately 1 to 100 days, with a prominent peak in the 5–10 day interval. This broad coverage ensures that the calibrating sample is representative of the entire period range expected for CCs.}\label{fig-period_distribution_DR3_based_signif_digit_3_SN.coeff_0_DCEP_F}
    \end{figure}

The periods of the reference sample
cover a broad interval from approximately 1 to 100 days, peaking between 5 and 10 days (see fig.~\ref{fig-period_distribution_DR3_based_signif_digit_3_SN.coeff_0_DCEP_F}). Such a comprehensive distribution ensures that our empirical relation remains representative of the full CC period range, accounting for the diverse physical properties of both short- and long-period pulsators.

\begin{figure}
   \centering
  \includegraphics[trim=0.0cm 1.0cm 0.0cm 2cm, clip, width=1.0\linewidth]{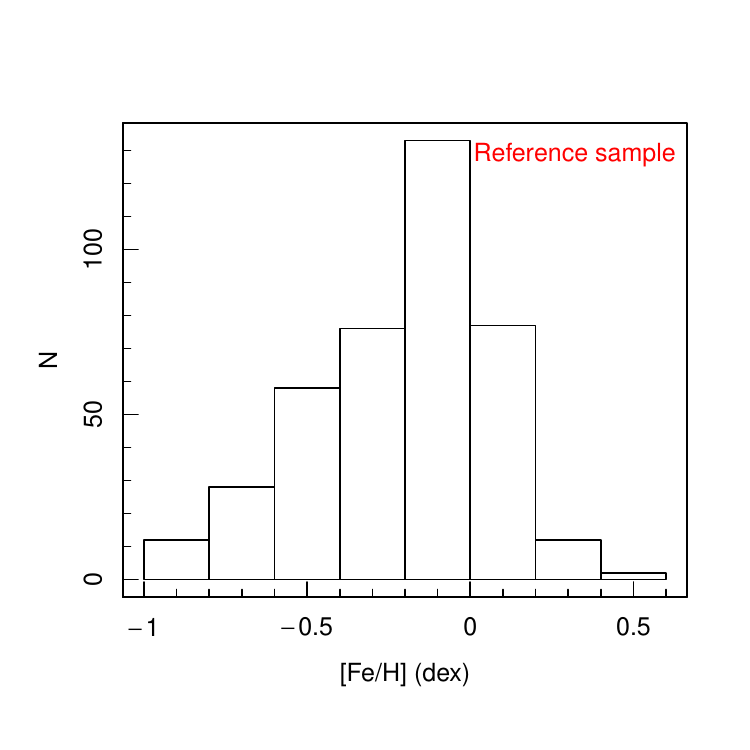}
   \caption{Distribution of spectroscopic metallicities for the reference sample of Galactic Classical Cepheids used in this work. The histogram shows the [Fe/H] values collected from the literature and adopted to calibrate the photometric metallicity relation.}\label{fig-reference_sample_fhe_distribution_DR3_based_signif_digit_3_SN.coeff_0_DCEP_F}
    \end{figure}

The metallicity distribution of the reference sample is shown in Fig.~\ref{fig-reference_sample_fhe_distribution_DR3_based_signif_digit_3_SN.coeff_0_DCEP_F}. The spectroscopic [Fe/H] values cover a broad range (-1.0 to 0.5 dex), ensuring that the calibration is representative of the metallicity regime expected for Galactic Classical Cepheids.

       \begin{figure}
   \centering
  \includegraphics[trim=0.1cm 1.0cm 1.0cm 2.0cm, clip, width=1.0\linewidth]{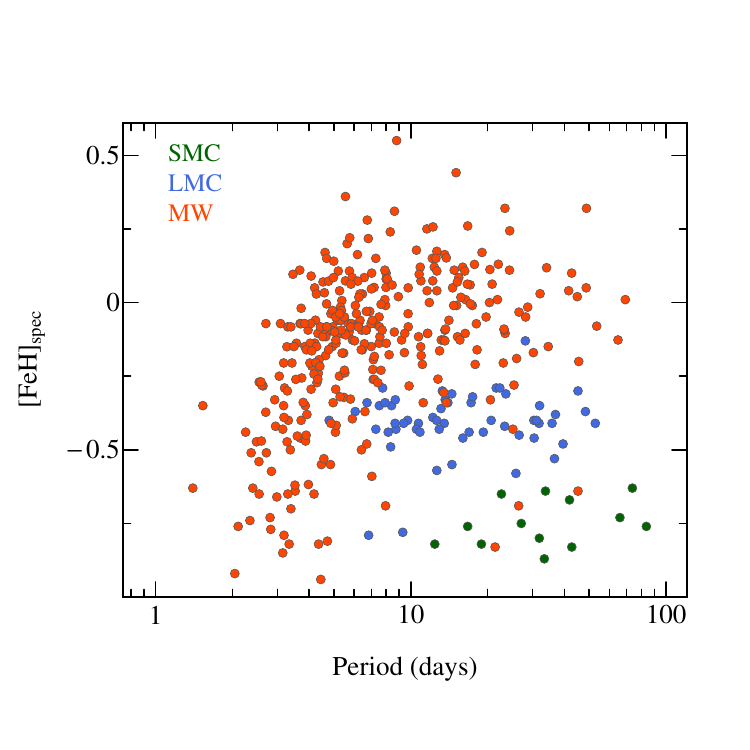}
   \caption{Spectroscopic iron abundance ([Fe/H]) as a function of the pulsation period (in log-scale) for the 398 Classical Cepheids in our calibration sample. The data points are color-coded by their parent stellar system: Milky Way (red), LMC (blue), and SMC (green). }\label{fig-feh_period_DR3_based_referee_signif_digit_3_SN.coeff_0_DCEP_F}
    \end{figure}

To better evaluate the combined coverage of our calibration sample, in Fig.~\ref{fig-feh_period_DR3_based_referee_signif_digit_3_SN.coeff_0_DCEP_F} we plot the spectroscopic [Fe/H] values against the pulsation period. The dataset is not perfectly uniform across the entire parameter space. In particular, Galactic Cepheids (MW) dominate the solar and metal-rich regimes but thin out considerably below $\rm [Fe/H] \sim -0.4$ dex, especially at longer periods. To mitigate this observational selection effect at low metallicities, we supplemented the calibrating sample with Cepheids from the LMC and SMC. While a lack of long-period, extremely metal-poor pulsators remains, the inclusion of the Magellanic Cloud targets ensures that the metal-poor regime is populated across a wide period range (from $\sim 4$ to over 90 days).

\section{[Fe/H] calibration}\label{sec-feh-fit}
Our primary goal is to establish a robust empirical relation between the spectroscopic [Fe/H] and the light curve characteristic parameters introduced above.
To this aim, we followed the subsequent steps: 

\begin{itemize}
    \item We first employed a RF algorithm to rank the set of available input parameters according to their relative importance\footnote{\label{note-rf}We utilized the \texttt{randomForest} package from the CRAN-R environment \citep{Breiman2001}. In this regression framework, the variable importance is quantified by using the Total Increase in Node Purity, defined as the cumulative reduction in the Residual Sum of Squares ($\rm RSS$) achieved by splitting on a given predictor. For a variable $X_j$, its overall importance $I(X_j)$ is computed across the entire forest as $I(X_j) = \frac{1}{N_{\rm tree}} \sum_{t=1}^{N_{\rm tree}} \sum_{\tau \in t: v(\tau) = X_j} \Delta \text{RSS}(\tau, X_j)$, where $\tau$ represents a node in tree $t$ where the split is driven by $v(\tau) = X_j$, and $\Delta \text{RSS}$ is the decrease in the residual variance resulting from that specific split.} in predicting the spectroscopic [Fe/H] values. We stress that the RF analysis was adopted solely for feature selection and not for metallicity prediction. Consequently, no explicit training-test split was performed, and the full sample was used to estimate the relative importance of the input parameters. Since the final metallicity relation is derived from an independent Ordinary Least Squares regression (see item below), the RF step does not constitute the predictive model itself but rather a procedure to reduce the dimensionality of the problem.
    \item Using only the  selected subset of highly important parameters, we proceeded to fit an Ordinary Least Squares linear model. To account for possible higher-order dependencies between the pulsation properties and metallicity, the final model includes not only the original parameters identified by the RF algorithm but also their powers up to the fourth order. The resulting linear regression model is defined as:
    \begin{equation}\label{eq:linear-relation}
        \rm [Fe/H] = \beta_0 + \sum_k \beta_k \cdot p_k
    \end{equation}
    where the terms $\rm p_k$ represent the full set of predictors, comprising both the selected Fourier parameters and their higher-order powers.
\end{itemize}
The latter step defines a simple analytical relation that can be directly used to estimate the [Fe/H] value of any fundamental mode CC, provided that its photometric light curve is characterised using a Fourier series model.

The following subsections describe in detail the implementation of these two steps, namely the RF-based feature selection and the subsequent regression analysis.   \begin{figure*}[!ht]
   \centering
  \includegraphics[trim=0.5cm 5cm 0.0cm 0cm, clip, width=1.0\linewidth]{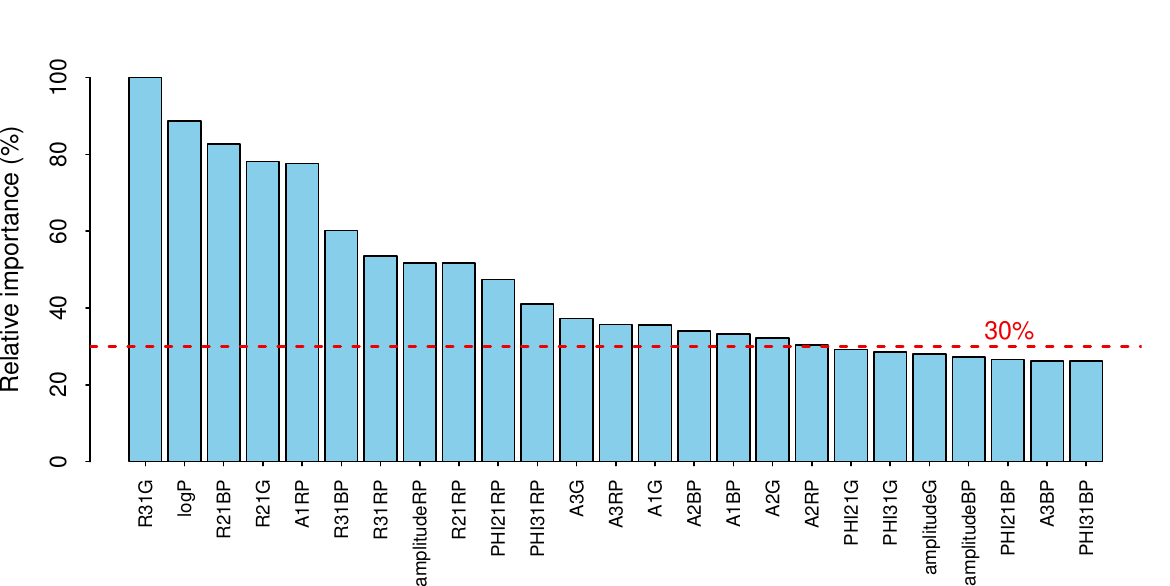}   \caption{Normalized relative importance (see note~\ref{note-rf}) of the input parameters derived using a RF algorithm, averaged over N = 500 realizations and displayed in descending order.  The dashed red line marks the empirical threshold adopted to select the final set of parameters. This threshold was identified using the elbow criterion, corresponding to the region where the importance distribution begins to flatten, indicating diminishing marginal gains in predictive power.}\label{fig-Dcep_F_FeH_RandomForest_Variable_Importance}
    \end{figure*}

\subsection{Feature Selection using Replicated Random Forest}
To identify the Fourier parameters carrying the most predictive power for $\rm [Fe/H]$, we employed the RF machine learning algorithm \citep{Breiman2001}. Given its stochastic nature, due to its random sub-sampling and feature selection at each node split, a single model run is often insufficient to guarantee a stable importance ranking. To ensure robustness, we implemented a replicated approach based on the following steps:
\begin{itemize}
    \item Model Definition: We defined a RF model using the generic formula $\rm [Fe/H]\sim \sum_{i=1}^{N_{par}} p_i$, setting the number of trees to $\rm N_t=500$ and enabling the calculation of variable importance vector $\bar{I}\equiv \{I_1, ...I_{N_{par}}\}$. Polynomial transformations were not included at this stage, as preliminary tests showed that the RF assigns comparable importance to each parameter and to its polynomial powers. This highlights a limitation of the RF method in discriminating among correlated features, such as parameters and their polynomial powers.
    \item Replication: The previous step for feature importance calculation was replicated $\rm N_{rep}=500$ times\footnote{This number was chosen as a compromise between result stability and computational time.}, parallelising the execution in order to significantly reduce the computational time. For each replicated run, we extracted the variable importance $N_{par}$-dimensional vector $\rm \bar{I}_j$ ($j=1,...N_{rep}$).
    \item  Averaging: The final, stable ranking was determined by calculating the mean importance for each parameter across all the $\rm N_{rep}$ replications:
    \begin{equation}
        \rm \bar{I}_{mean} = \frac{1}{N_{rep}}\sum_{j=1}^{N_{rep}}\bar{I}_j
    \end{equation}
\end{itemize}

To select a subset of highly influential parameters, we first normalised the components of the mean importance vector ($\rm \bar{I}_{mean}$) by scaling them to the maximum observed importance, thus transforming the scores into a percentage. Next, we constructed the ranked importance plot (see Fig.~\ref{fig-Dcep_F_FeH_RandomForest_Variable_Importance}), which displays the parameters in descending order of their normalised contribution. 

To establish an objective baseline for feature selection, we applied the \textit{Kneedle} detection algorithm \citep{sat11} to the relative importance distribution, which analytically isolated the point of maximum curvature (the "elbow" of the curve) at a threshold of $37.2\%$.

However, a selection strictly based on this $37.2\%$ cutoff results in a loss of information necessary to resolve low-metallicity regimes. A preliminary consistency check on the final $\rm [Fe/H]$ distributions, evaluated using our external validation samples of Galactic, LMC, and SMC Cepheids (see sec.~\ref{sec-application}), revealed that this strict mathematical threshold leads to systematically offset and uncentered predictions, particularly for the metal-poor SMC system. 

To resolve this issue, we systematically explored lower importance thresholds, finding that a 30\% cutoff represents the best compromise between accurately retrieving the expected spectroscopic peaks of the MW, LMC, and SMC frameworks and mitigating the risk of overfitting from an excessive number of parameters. A comprehensive analysis of these galactic distributions, along with a detailed discussion of the validation results and the associated figures, is presented and fully expanded in Sec. ~\ref{sec-application}.

\subsection{The best fit equation}
Adopting only the variables identified as most important by the RF analysis, we constructed the linear relation introduced in Eq.~\ref{eq:linear-relation} through a two-step procedure. First, an initial Ordinary Least Squares (OLS) fit including all selected variables and their polynomial terms (up to fourth order) was performed  to identify the statistically significant\footnote{In this context, we define as statistically significant those coefficients that differ from zero by more than their formal uncertainty.}  coefficients. Subsequently, a second fit was carried out, retaining only the variables found to be significant in the first step, yielding the final analytical relation for [Fe/H].
The best combination of parameters we obtained is the following:
\begin{align}
[\mathrm{Fe/H}] =\; & \beta_0 
+ \beta_1 \left(R_{31}^{G}\right)^{4}
+ \beta_2 \left(R_{31}^{G}\right)^{2}
+ \beta_3 \left(A_{1}^{G}\right)^{4}
+ \beta_4 \left(A_{1}^{G}\right)^{2} \notag \\
& + \beta_5 \left(R_{21}^{RP}\right)^{2} 
+ \beta_6 \left(R_{21}^{RP}\right)
+ \beta_7 \left(A_{1}^{RP}\right)^{2}
+ \beta_8 \left(A_{2}^{RP}\right)^{2} \notag \\
& + \beta_9 \left(A_{2}^{RP}\right)
+ \beta_{10} \left(A_{3}^{RP}\right)^{4}
\label{eq:dcepf-feh}
\end{align}
where $A^X_i$s represents the amplitude of the i-th component of the fitted Fourier truncated series and $\rm R^X_{i1}$ is defined in Eq.~\ref{eq-fourier-params}.
This relation is characterised by an rms of the residuals equal to 0.213 dex and a Pearson correlation index of  0.61. The coefficients $\beta_k$ are listed in Table~\ref{tab:feh_coeff}.

For completeness, and to explicitly account for the widely different scales of uncertainty between first-order and higher-order Fourier parameters, we also tested an alternative calibration based on a Weighted Least Squares (WLS) scheme. Under this framework, some parameters that were statistically significant in the unweighted OLS fit became non-significant, leading to a different subset of predictors. Although this weighted model successfully reproduces the expected peak metallicity distributions for the MW, LMC, and SMC systems, it yields a significantly lower correlation coefficient (0.40 vs 0.61) and a larger dispersion (rms of residuals is equal to 0.336 dex vs 0.213 dex) compared to our primary relation.  We therefore retain the unweighted OLS model as our best calibration, while providing the WLS relation and coefficients in Appendix ~\ref{app:coefficients}.

\begin{figure*}[!ht]
   \centering
  \includegraphics[trim=0.0cm 0cm 0.0cm 0cm, clip, width=1.0\linewidth]{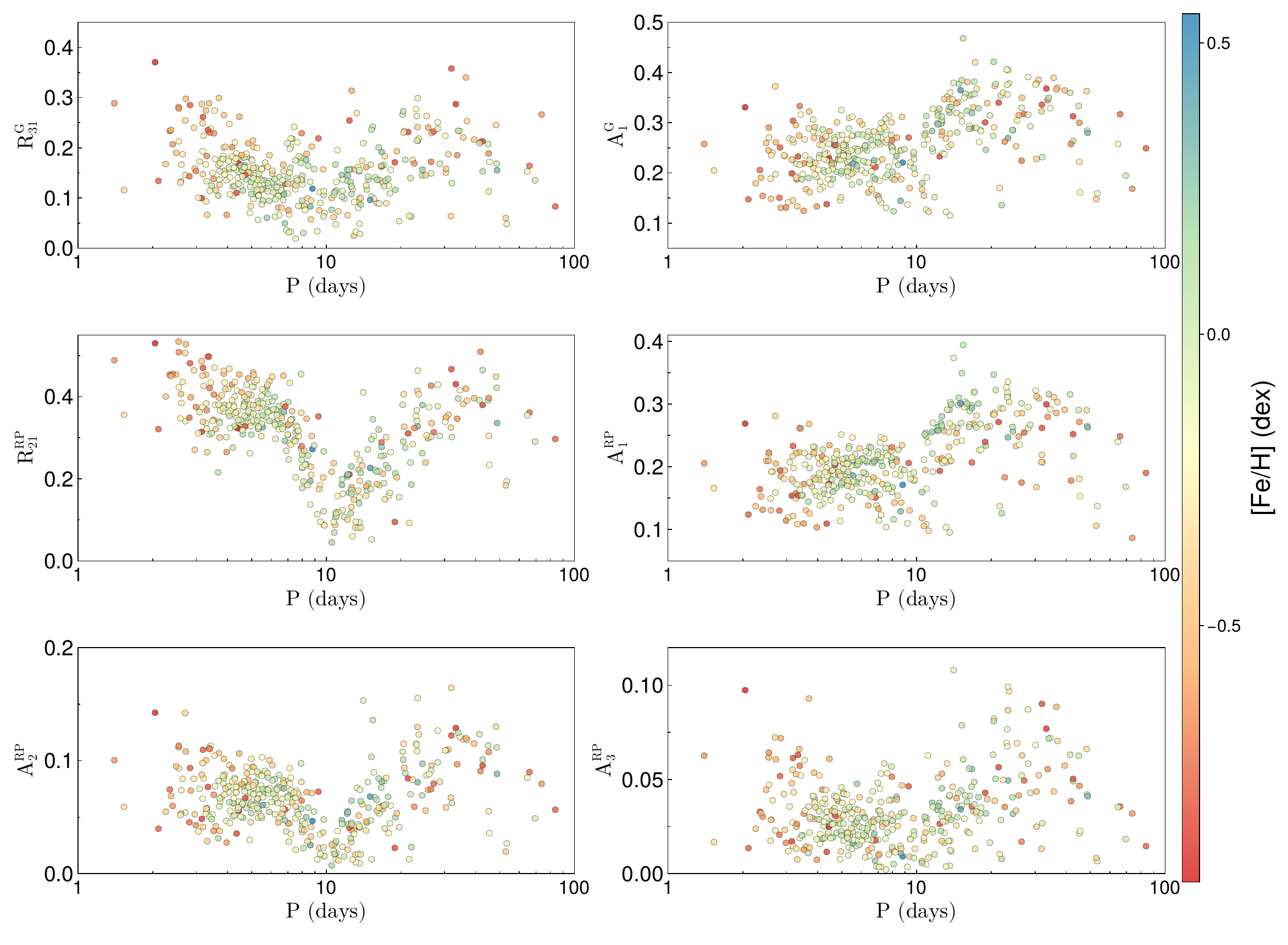}   \caption{Distribution of the Gaia DR3 light curve Fourier parameters used in the final calibration relation (Eq.~\ref{eq:dcepf-feh}) as a function of the log-period (logP) for the calibrating Cepheid sample. The panels display (from top-left to bottom-right): $\rm R_{31}^G$, $\rm A_{1}^G$, $\rm R_{21}^{RP}$, $\rm A_1^{RP}$, $\rm A_2^{RP}$, and $\rm A_3^{RP}$. The data points are color-coded by their spectroscopic iron abundance [Fe/H]. The structural variations associated with the Hertzsprung progression around P$\simeq$10 days are clearly visible, along with a secondary stratification in metallicity at fixed period values.}\label{fig-fou_par_zcol_ref_sample}
    \end{figure*}

To provide a visualization of the calibrating sample and to justify the multi-band regression model, in Fig.~\ref{fig-fou_par_zcol_ref_sample} we show the distribution of the six selected Fourier parameters entering Eq. ~\ref{eq:dcepf-feh} as a function of the pulsation period. The data points are color-coded according to their spectroscopic [Fe/H] values. The complete spectroscopic and photometric properties of this calibrating dataset, including light curve quality indicators and individual uncertainties for each parameter, are listed in Table ~\ref{tab:ref_sample_param} in Appendix ~\ref{app:Reference sample}.

As expected, the overall morphology in all panels follows the well-known Hertzsprung progression. The effect is especially evident in the amplitude ratios ($\rm R_{31}^G$ and $\rm R_{21}^{RP}$) and the amplitudes $\rm A_2^{RP}$ and $\rm A_3^{RP}$, which display the characteristic sharp V-shape transition around P $\sim$ 10 days \citep{mar24} that corresponds to the Hertzsprung progression center.

We note that the pulsation period does not explicitly enter Eq.~\ref{eq:dcepf-feh} as an independent variable. This reflects the fact that the metallicity information carried by the pulsation period is already encoded in the Fourier parameters adopted in the fit. 

Moreover, no term involving the BP band Fourier parameters appears in the final relation. This indicates that, within the adopted fitting procedure, the BP observables do not provide independent constraining power on [Fe/H] beyond what is already captured by the G and RP band parameters.

Finally, no phase-difference Fourier parameters (i.e. $\phi_{i1}$ terms defined in eq.~\ref{eq-fourier-params}) enter the final relation. Although these quantities were initially considered among the candidate predictors, their coefficients were not found to be statistically significant in the fitting procedure. This suggests that, in the present sample, the phase-difference parameters do not provide additional independent information on the iron abundance beyond that already contained in the amplitude-based descriptors.

It is worth noting that the Fourier parameters used in the calibration 
depend not only on metallicity but also on other stellar properties such as mass, luminosity, effective temperature, and convection efficiency \citep[see, e.g.,][and references therein]{mar24}. In addition, the well-known Hertzsprung progression introduces systematic variations of the light-curve shape with pulsation period. Pulsation models show that these effects can lead to partial degeneracies among the parameters governing the light-curve morphology (see, e.g., \citealt{bha17,mar24}). In this context, the empirical calibration presented here should be regarded as exploiting the statistical correlations present in the observed sample rather than isolating a purely metallicity-driven dependence of the Fourier parameters.

\subsection{Performance on the Reference Sample} 
To evaluate the internal consistency of the calibrated relation, we first performed a direct comparison between the photometrically predicted abundances ([Fe/H]$\rm _{phot}$) and the high-resolution spectroscopic values ([Fe/H]$\rm _{spec}$) for the reference sample

\begin{figure*}
   \centering
  \includegraphics[trim=0cm 6.0cm 1.5cm 0cm, clip, width=1.0\linewidth]{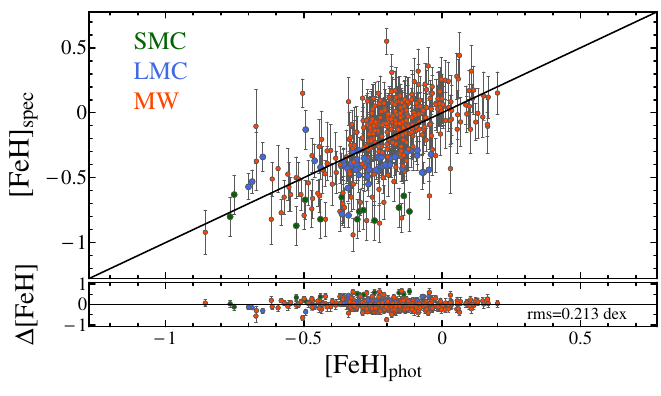}
   \caption{Comparison between the spectroscopic abundances ([Fe/H]$_{\rm spec}$) and the photometrically derived metallicities ([Fe/H]$_{\rm phot}$) for the Cepheids of the reference sample. Data points are color-coded according to their host system: Milky Way (red), LMC (blue), and SMC (green), with the solid black line representing the 1:1 identity relation. The bottom panel shows the residuals ($\Delta$[Fe/H] = [Fe/H]$_{\rm spec}$ - [Fe/H]$_{\rm phot}$) as a function of the photometric metallicity, yielding an overall dispersion of ${\rm rms} = 0.213$ dex. }\label{mw_specFeh_vs_photFeh_comparison_DR3_based_referee_signif_digit_3_SN.coeff_0_DCEP_F}
    \end{figure*}

As shown in Fig.~\ref{mw_specFeh_vs_photFeh_comparison_DR3_based_referee_signif_digit_3_SN.coeff_0_DCEP_F}, the photometric estimates are in overall good agreement with the spectroscopic measurements for the Milky Way and LMC samples, with data points distributed along the identity line. However, the comparison also highlights a visible departure from the 1:1 relation in the metal-poor regime populated by the SMC Cepheids, where the photometric relation tends to systematically overestimate the metallicity. This trend confirms that the low-metallicity framework remains a critical regime, although its impact could be significantly mitigated by expanding the available sample of spectroscopic measurements at very low abundances \citep{rip21, rip22, rip23, rip26, tre24a, tre24b, tre26}.

Despite the residual tension seen for the SMC, the global residuals ($\rm \Delta[Fe/H] = [Fe/H]_{spec} - [Fe/H]_{phot}$), displayed in the bottom panel, yield a negligible overall median offset of $-0.014$ dex and a global dispersion of $0.213$ dex. This scatter, while non-negligible, is consistent with the typical uncertainties associated with both the Fourier parameter determination and the underlying spectroscopic measurements. Despite the highlighted limitations at low abundances, this internal comparison gives a reasonable degree of confidence for applying the relation to external datasets.

\section{Testing the relation to MW, LMC, SMC Cepheids}\label{sec-application}
As anticipated in Sect.~\ref{sec-feh-fit}, the choice of our optimal importance threshold was inherently tied to the model's performance across different galactic environments. To fully explore and validate the capabilities of our best-fit relation (Eq.~\ref{eq:dcepf-feh}), we applied it to three distinct validation samples of fundamental-mode CCs located in the MW, LMC, and SMC, respectively. This comprehensive environmental analysis allows us to test the reliability and robustness of the derived metallicity relation under different metallicity regimes.

We selected CCs from the Gaia DR3 catalogue and retained only fundamental mode pulsators, yielding a sample of 7712 sources. For each of these objects, we constructed a Fourier model of the light curve in the G, $\rm G_{BP}$ and $\rm G_{RP}$ bands, as described in Sect.~\ref{sec:fourier-analysis}.
This sample was refined through a two-step filtering process. First, we excluded sources for which the Fourier decomposition of the light curves did not provide the necessary higher-order coefficients. Specifically, several sources possessed fits truncated at an order lower than that required by our relation (e.g., lacking parameters such as $A_3^G$). Second, we computed [Fe/H] by applying Eq.~\ref{eq:dcepf-feh} and discarded objects yielding values outside the range $\rm -1.5 < [Fe/H] < 0.6$ dex. This interval, although slightly broader than the metallicity range covered by the reference sample, reflects the range supported by spectroscopic measurements of CCs in the MW and nearby galaxies \citep[see e.g.][]{rom08, lem13, das22, rom22, tre24b, rip25}, and avoids excluding sources with plausible metallicities at the edges of the distribution. We note that only a very small fraction of the initial sample (about 0.6\%) falls outside this interval, indicating that the derived relation provides physically meaningful metallicity estimates for the vast majority of the sources.

A galaxy flag (MW, LMC, or SMC) was assigned to each source based on its position on the sky, adopting the spatial subregions defined by \citet[][see their Table~1]{rip23}. The initial sample also contains 472 CCs located in the M31 and M33 sky regions. These were excluded from the analysis because their light-curve models typically include fewer than three harmonics, which prevents the computation of the Fourier parameters required by the adopted [Fe/H] relation.

After applying these criteria, the final sample used for this test is defined as follows:
\begin{itemize}
\item \textit{SMC sample:} Of the 2485 initial candidates, 761 were removed due to an insufficient number of Fourier harmonics, and 21 were excluded because their metallicity values lie outside the adopted [Fe/H] range, yielding a final sample of 1703 sources.
\item \textit{LMC sample:} From an initial set of 2355 stars, 629 were excluded due to an insufficient number of Fourier harmonics, and 1 was removed for falling outside the adopted metallicity range, resulting in a final sample of 1725 sources.
\item \textit{Galactic sample:} Out of 2400 initial sources, 719 were excluded due to an insufficient number of Fourier harmonics, and 7 were discarded because their metallicity values fell outside the adopted [Fe/H] range, yielding a final sample of 1674 sources.
\end{itemize}

A small fraction of targets was discarded a posteriori because their predicted metallicities fell outside the adopted boundaries. This discrepancy is entirely driven by the local photometric quality of the light curves. Specifically, we checked the quality flag of the Fourier fit, defined as the ratio between the standard deviation of the raw photometric data (column 19-20-21 of tab.~\ref{tab:ref_sample_param}) and the rms of the residuals (column 16-17-18 of tab.~\ref{tab:ref_sample_param}).  We found that for nearly all the discrepant sources, this parameter drops below 7.0 in the $\rm G_{RP}$ band.
Below this threshold, the Fourier decomposition starts to fail to accurately reproduce the light curve morphology, thus introducing significant noise into the higher-order amplitudes and amplitude ratios used by our empirical relation.

\begin{figure}
   \centering
  \includegraphics[trim=0cm 0.0cm 6.0cm 0cm, clip, width=1.0\linewidth]{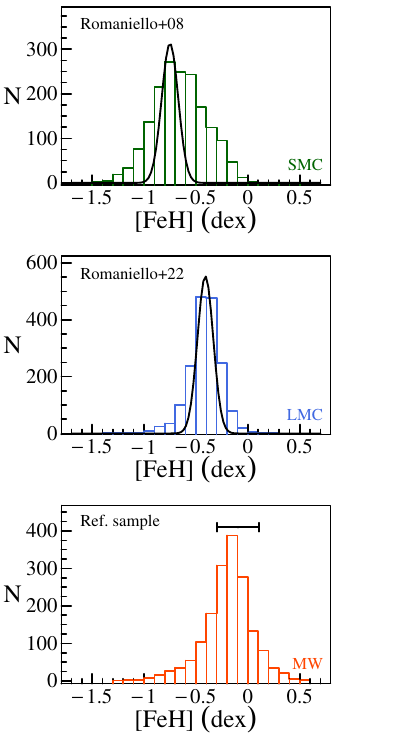}
   \caption{Distributions of the photometrically predicted $[Fe/H]$ values obtained by applying Eq. (5) to our validation samples of fundamental-mode Classical Cepheids across different galactic systems. Panels show, from top to bottom, the results for the SMC (green histogram), the LMC (blue histogram), and the Milky Way (red histogram). For the Magellanic Clouds (top and middle panels), the smooth black curve represents a reference Gaussian centered on the average spectroscopic metallicity from the literature, with a width  corresponding to the error on the mean value published by \citet{rom08} for the SMC (14 stars)  and \citet{rom22} for the LMC (88 stars). To allow an immediate comparison of the distributions without compressing the vertical scale of the histograms, these reference Gaussian curves have been scaled to match the peak height of each photometric distribution. For the Milky Way (bottom panel), the black horizontal bar denotes the median metallicity and the dispersion estimated from the (scaled) MAD of the MW reference sample.}\label{fig-feh_distributions_DCEP_F}
    \end{figure}

Fig.~\ref{fig-feh_distributions_DCEP_F} shows the distributions of the predicted metallicity for the fundamental mode CCs in the MW, LMC, and SMC samples. The performance of our calibration is validated by comparing the resulting [Fe/H] distributions with independent spectroscopic reference values from the literature. For the Magellanic Clouds (top and middle panels), these references are visualized as full Gaussian curves centered on the literature mean and with a width corresponding to the nominal spectroscopic population dispersion. To facilitate a direct morphological comparison, these curves are scaled to match the peak height of each photometric distribution. For the MW (bottom panel), the literature reference is denoted by a black horizontal bar representing the median metallicity and its associated dispersion.
The metallicity distributions exhibit a clear trend following the expected chemical enrichment of the three galaxies. Specifically:
\begin{itemize}
\item SMC (top panel): The selected sample (green histogram) shows a distribution peaked around -0.75 dex, matching the central value of the spectroscopic reference ($-0.75 \pm 0.08$ dex) from \citet{rom08}. However, as visually highlighted by the comparison with the narrower reference curve, the photometrically predicted metallicities for the SMC exhibit a considerably broader dispersion, with a scaled MAD\footnote{The Median Absolute Deviation (MAD) is defined as $\text{median}(|x_i - \text{median}(x)|)$. For comparison with the standard deviation, we utilize the scaled MAD, $\sigma_{\text{MAD}} = 1.4826 \times \text{MAD}$, which is statistically equivalent to $\sigma$ for a normal distribution.} of $0.25$ dex, tailing up to solar abundances and down to $\sim -1.5$ dex. Rather than reflecting an intrinsic chemical property of the SMC, this large spread primarily underscores the current shortcomings of the photometric calibration process in the very metal-poor domain. Because the empirical relation lacks a statistically large and uniform training sample below $\rm [Fe/H] \sim -0.5$ dex, the regression framework becomes less constrained. This emphasizes that while the relation remains a valuable statistical tool, its individual predictions at low metallicities must be treated with caution until more metal-poor calibrators can be integrated. 
\item LMC (middle panel): The distribution for the photometric [Fe/H] values (blue histogram) is centred at -0.41 dex with a (scaled) MAD equal to 0.13 dex. This result shows a good agreement with the central peak of the reference spectroscopic distribution ($-0.409 \pm 0.076$ dex) from \citet{rom22}, although our photometric distribution still displays slightly broader wings than the nominal literature Gaussian. The overall distribution appears symmetric and well constrained, confirming the general consistency of our relation at sub-solar metallicities.  
    \item MW (bottom panel): The predicted [Fe/H] values for the selected sample are peaked around -0.15 dex, with a (scaled) MAD of 0.23 dex, showing good agreement with the range provided by the Galactic sources taken from our reference sample, which has a median value equal to -0.09 dex and a (scaled) MAD of 0.20 dex. The distribution is broader compared to the Magellanic Clouds, reflecting the more complex chemical gradient of the Galactic disk. 
\end{itemize}

    Overall, the progressive shift of the peaks toward more metal poor values as we move from the MW to the LMC and SMC demonstrates that our Fourier parameter based relation effectively captures  metallicity variations across different galactic environments.    
    
\begin{figure}
\includegraphics[trim=0cm 0.0cm 0.0cm 0cm, clip, width=1.0\linewidth]{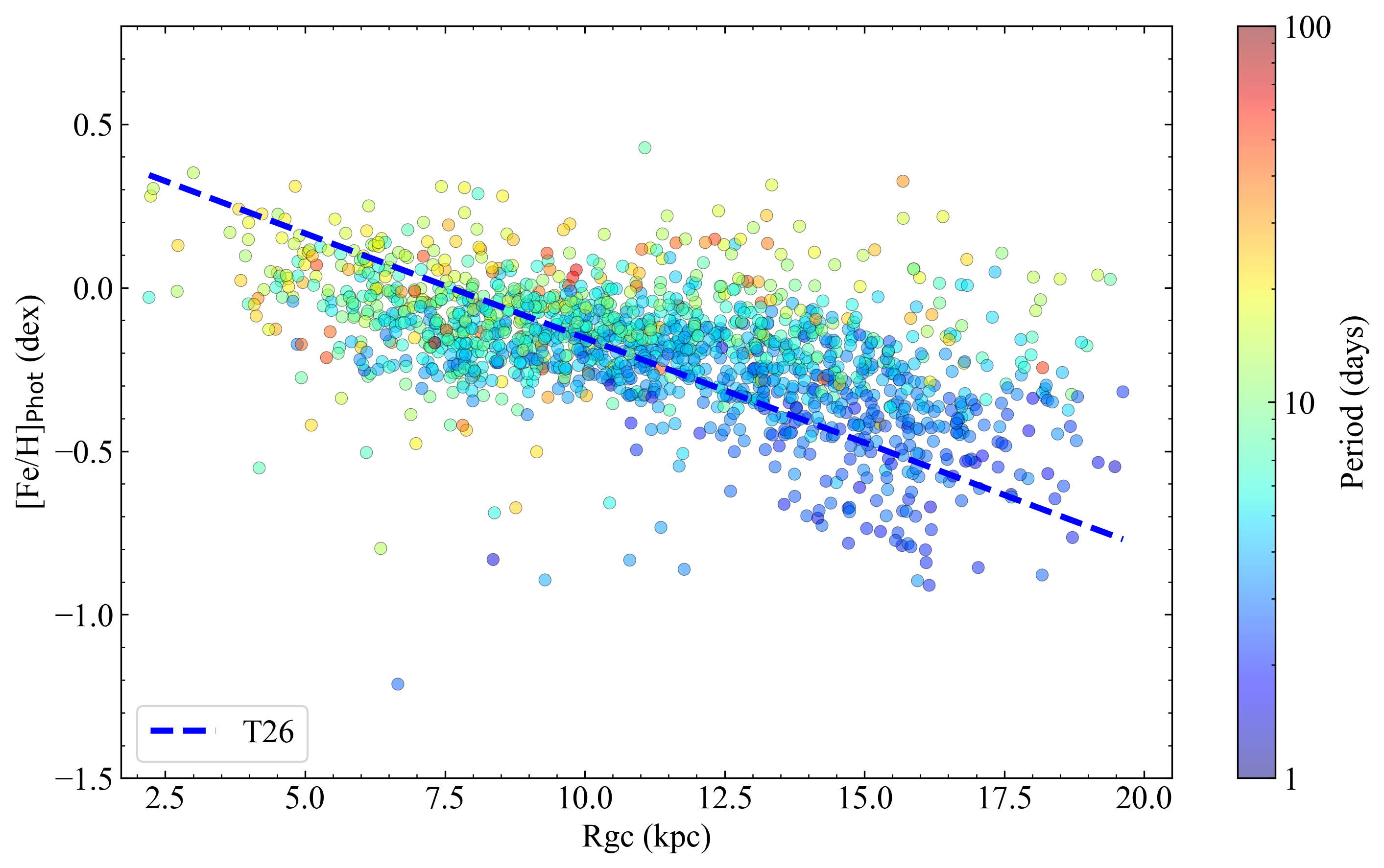}
   \caption{Photometric metallicity ([Fe/H]$_{\rm Phot}$) as a function of galactocentric distance ($R_{\rm gc}$) for the Galactic Classical Cepheids in our test sample. Data points are color-coded according to their pulsation period. The dashed blue line represents the linear metallicity gradient derived by \citet[][T26]{tre24a}. }\label{fig-Reproduced_Metallicity_Gradient_AllModels}
    \end{figure}

As a further independent check, we investigated the capability of our relation to reproduce the metallicity gradient of the Galactic disk. In Fig.~\ref{fig-Reproduced_Metallicity_Gradient_AllModels}, we plot the photometrically derived metallicities against the galactocentric distance for our sample of Galactic Cepheids, with data points color-coded by their pulsation period. 
Our photometrically inferred metallicities exhibit a clear, negative trend with increasing galactocentric radius. We compared our distribution with the recent linear spectroscopic gradient derived by \citet{tre24a}, overlaid as a dashed blue line. The comparison reveals a good overall agreement in terms of the absolute metallicity scale, particularly at solar radii, although our photometrically derived gradient appears visibly shallower than the spectroscopic one obtained by \citet{tre24a}. While our relation successfully captures the macroscopic decreasing trend from the inner to the outer disk over a wide baseline ($R_{\rm gc} \sim 2.5$ to $20$ kpc), it exhibits a mild flattening at large galactocentric distances.

Finally, we note that two potential sources of systematic bias should be considered as caveats when applying our broad-band relation to independent datasets: phase-dependent extinction and stellar blending. 
Because of the Gaia $G$ band is exceptionally wide, the effective extinction depends on the star's instantaneous spectral energy distribution \citep[e.g.,][]{dan18}. As a Cepheid pulsates, its temperature variations modulate the total extinction in the $G$ band between maximum and minimum light, which could theoretically distort the observed light-curve shape and the derived Fourier parameters. Similarly, stellar blending within the Gaia selection window introduces a constant flux contamination that systematically dampens the pulsation amplitudes, directly affecting the photometric metallicity estimates.

While a detailed quantitative modeling of these coupled phenomena goes beyond the scope of this work, they represent higher-order effects that can introduce biases. We therefore caution that when using this empirical relation, users should carefully account for the specific sky region of the targeted sources, avoiding areas characterized by extreme stellar crowding or highly non-uniform interstellar reddening.

\subsection{Comparison with previous photometric calibrations}
\label{subsec:lit_comparison}
\begin{figure*}
    \centering
    \includegraphics[width=\textwidth]{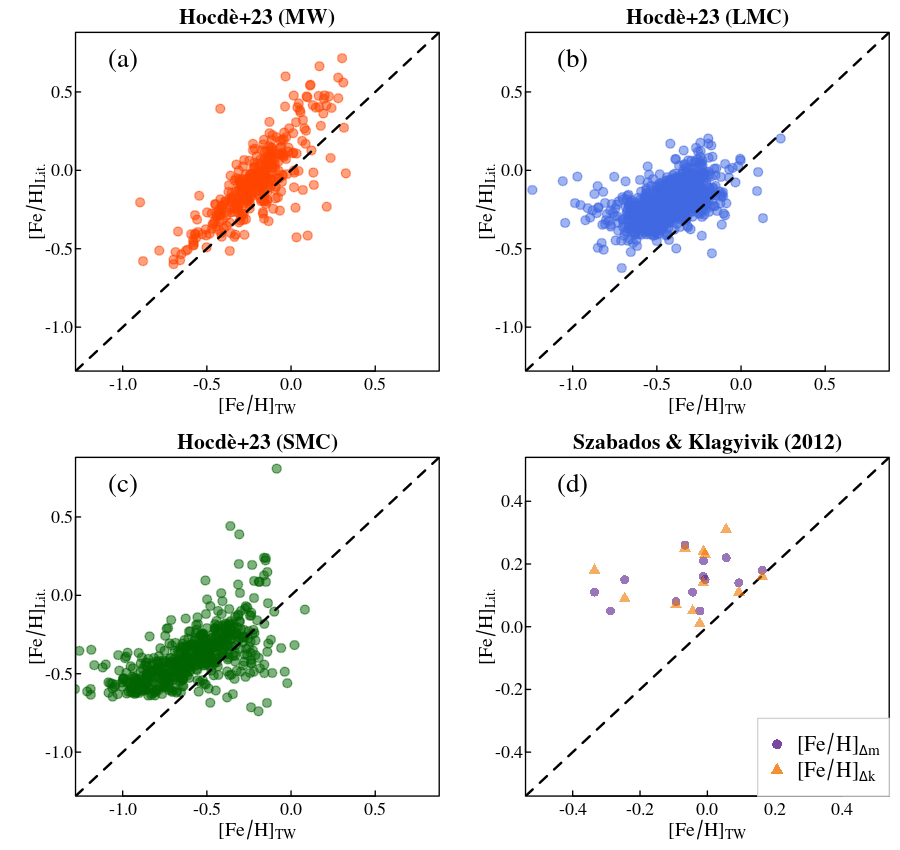}
    \caption{Comparison between the photometrically predicted metallicities of this work on the $x$-axis ($\rm [Fe/H]_{TW}$) and previous calibrations in the literature on the $y$-axis ($\rm [Fe/H]_{Lit.}$). Panels (a), (b), and (c) display the comparison with the values obtained by \citet{hoc23} using the $I$ photometric band for the MW, LMC, and SMC, respectively. Panel (d) shows instead a comparison between our values and those from \citet{sza12} obtained for a sample of Galactic Cepheids through two different photometric relations (see their Tab.~4). The black dashed line in each panel represents the 1:1 identity line.}
    \label{fig:lit_comparison}
\end{figure*}

To further assess the predictive performance and robustness of our calibration, we performed a quantitative comparison with alternative photometric metallicity indicators from the literature. First, we performed a coordinate cross-match between our independent dataset and the sample published by \citet{hoc23}. This procedure yielded 369, 1265, and 572 classical Cepheids in common between the two works for the MW, LMC, and SMC, respectively. \citet{hoc23} derived the photometric metallicities of these variables using an empirical relation calibrated in the $I$ band. The resulting quantitative comparison is presented in panels (a), (b), and (c) of Fig.~\ref{fig:lit_comparison}. An additional comparison was carried out with the work of \citet{sza12}, in which the authors proposed two separate photometric metallicity relations ($\rm [Fe/H]_{dm}$ and $\rm [Fe/H]_{dk}$) based on the wavelength dependence of multi-color light-curve amplitudes. This comparison is displayed in panel (d) of the same figure.
The comparison with \citet{hoc23} clearly highlights the limitations affecting their single-band linear relation. In the LMC panel (b), their metallicities exhibit a pronounced horizontal flattening  around $\rm [Fe/H]_{Hocde} \sim -0.2$~dex, an effect that becomes even more critical in the low-metallicity regime of the SMC shown in panel (c). While our model successfully tracks the metal-poor stars continuously down to $\rm [Fe/H] \sim -1.0$~dex, the single-band estimates fail to recover values below $-0.6$~dex,  compressing the distribution of the obtained [Fe/H] values, as explicitly acknowledged by the authors themselves.

Similarly, the comparison with \citet{sza12} in panel (d) reveals an apparent horizontal distribution that closely resembles the flattening observed in the \citet{hoc23} panels. Notably, the photometric $\rm [Fe/H]$ estimates from \citet{sza12} are strictly confined to positive values and never reach the sub-solar regime covered by our predictions. However, this discrepancy is inherently a consequence of the underlying calibration samples; the reference spectroscopic compilation used by \citet{sza12} simply does not contain negative values for these stars, while the stars in common between our and their sample (16 Cepheids) have both negative and positive values spanning from -0.1 dex up to +0.1 dex.

Analogously, a similar baseline mismatch explains the systematic zero-point offset visible in the \citet{hoc23} comparisons, where the matched populations lie systematically above the 1:1 identity line. To quantify this effect, we cross-matched the sources in common between the respective spectroscopic training samples and directly compared their reference spectroscopic abundances. This analysis revealed that our reference spectroscopic values are systematically lower by approximately $0.15$~dex compared to those adopted by \citet{hoc23}. This fundamental shift between the two foundational spectroscopic scales propagates directly into the final photometric predictions, naturally generating the observed systematic offset along the $y$-axis.

Finally, these tests hint the physical advantages of moving from narrow or single-passband calibrations to the simultaneous multi-band formulation offered by the Gaia filters.

\section{Conclusions}\label{sec-conclusions}
In this work, we present and validate a new photometric relation to estimate the metallicity ([Fe/H]) of fundamental mode CCs based on the Fourier parameters in the Gaia bands. Our analysis and the subsequent testing on different galactic environments lead to the following conclusions:
\begin{itemize}
    \item The RF method was used to identify the most significant features for predicting [Fe/H] among the available Fourier parameters. This machine learning feature selection allowed us to identify the parameters with the highest predictive power. Based on these results, we derived a linear relation that incorporates the selected parameters and their powers up to the fourth order.
        \item A direct star-by-star comparison with the reference sample of 398  CCs with high resolution spectroscopic data reveals a negligible systematic bias, with an average offset of $-0.017$ dex, while the dispersion of the residuals is 0.20 dex. Our relation is highly effective for identifying the central metallicity value of a stellar population.
        \item We successfully applied the derived relation to large validation samples of Classical Cepheids in the MW, LMC, and SMC. The resulting photometric [Fe/H] distributions correctly reproduce the progressive shift of the peaks following the expected chemical enrichment of these galaxies. While the global peak values show a good agreement with independent literature measurements, the SMC distribution exhibits an artificially broad dispersion. As discussed in Sec.~\ref{sec-application}, this large spread represents a current limitation of the photometric calibration due to the undersampling of spectroscopic calibrators in the very metal-poor domain ([Fe/H] < -0.5 dex), rather than an intrinsic chemical property of the SMC.
\item A key aspect of the present calibration is that it combines the multi-band information contained in the Gaia light curves. By exploiting Fourier parameters derived from multiple Gaia bands, the relation captures a richer description of the pulsation morphology than approaches based on a single band \citep[see e.g.][]{sza12, hoc23}, likely contributing to the improved precision of the resulting photometric metallicities. 
\item While our wide-band approach is susceptible to second-order effects like phase-dependent reddening, that makes the effective extinction coefficient modulates during the pulsation cycle as the star's temperature changes and flux blending in crowded regions, which systematically dampens the observed amplitudes and can skew individual photometric metallicity estimates, direct cross-matches verify its superior continuity down to $\rm [Fe/H] \sim -1.0$~dex.
\end{itemize}

The relation derived in this study represents a powerful and efficient tool for investigating the chemical properties of distant stellar systems (e.g., external galaxies) where Gaia light curves are available for a sufficient number of CCs, but high-resolution spectroscopy remains unfeasible. We emphasize that this relation is not intended as a unique or definitive statistical framework for photometric [Fe/H] calibrations. Our goal is strictly practical: to provide a robust, physically validated tool ready for immediate use by the community. While alternative statistical pipelines or different parameter selections could certainly be explored, the formula presented here offers a highly effective asset for mapping large-scale metallicity variations. This practical utility will be crucial for rapidly exploiting upcoming datasets, beginning with the next Gaia Data Release 4 (DR4). Looking further ahead, this same photometric approach can be naturally extended to future wide-field surveys, such as the Vera C. Rubin Observatory (LSST), opening up the possibility of mapping classical Cepheid metallicities across significantly larger cosmic volumes.

\begin{acknowledgements}
The authors thank the anonymous Referee for his valuable suggestions that greatly improved the robustness and completeness of this work.
      This research has used the SIMBAD database operated at CDS, Strasbourg, France ad data from the European Space Agency (ESA) mission Gaia, processed by the Gaia Data Processing and Analysis Consortium. Funding for the DPAC has been provided by national institutions, in particular, the institutions participating in the Gaia Multilateral Agreement.  We acknowledge funding from: INAF GO-GTO grant 2023 “C-MetaLL - Cepheid metallicity in the Leavitt law” (P.I. V. Ripepi); PRIN MUR 2022 project (code 2022ARWP9C) 'Early Formation and Evolution of Bulge and Halo (EFEBHO),' PI: Marconi, M., funded by the European Union – Next Generation EU; Large Grant INAF 2023 MOVIE (P.I. M. Marconi).  T.S. and G.D.S. thank the Istituto Nazionale di Fisica Nucleare (INFN), Naples section, for support through specific initiatives QGSKY. G.D.S. thanks Gaia DPAC funds from INAF and the Agenzia Spaziale Italiana (ASI) (2025-10-HH.0 PI: M.G. Lattanzi). Finally, this paper is based on work supported by COST Action CA21136, "Addressing Observational Tensions in Cosmology with Systematics and Fundamental Physics (CosmoVerse)," funded by COST (European Cooperation in Science and Technology).
\end{acknowledgements}


\appendix

\section{Homogeneisation of Galactic CCs metallicities}
\label{sec:metallicity_fit}

To obtain a sample of CC metallicites as homogeneous as possible, we assumed the abundances by \citet{tre24b} as reference and used the stars in common between this work and \citet{kov22} or \citet{das22} to calculate linear relations to put these works abundances in the \citet{tre24b} scale. To this aim, we employed the Orthogonal Distance Regression (ODR) technique to determine the linear relationship between the datasets. Unlike standard Ordinary Least Squares (OLS), which only accounts for errors in the dependent variable, ODR accounts for uncertainties in both the independent and dependent variables, providing a more robust and unbiased estimate of the functional relationship \citep{Boggs1990}.

\begin{figure}[htbp]
    \centering
    \includegraphics[width=9cm]{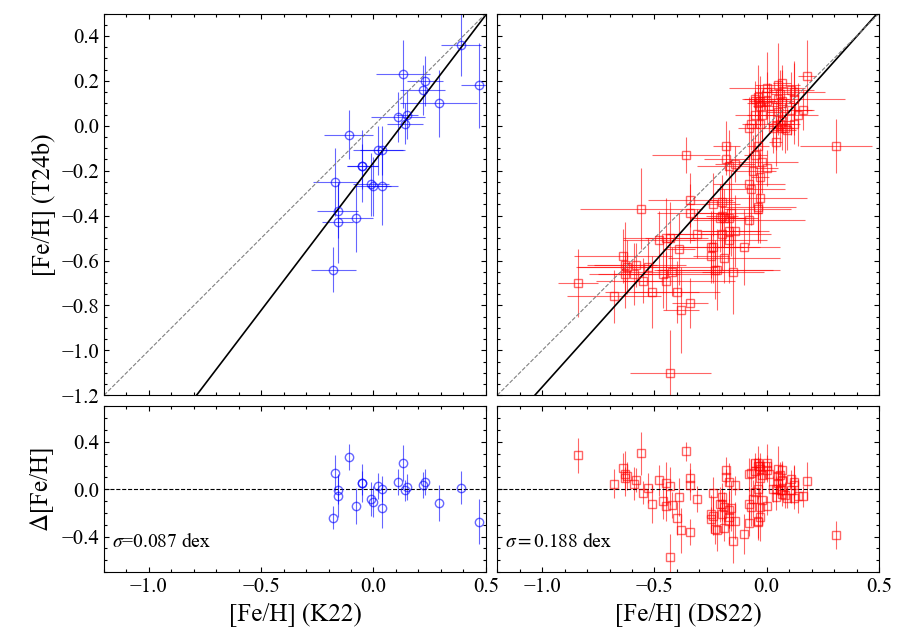}
    \caption{Comparison of derived [Fe/H] values with \citet{kov22} (left panels) and \citet{das22} (right panels). The top row shows the data points with $1\sigma$ error bars and the best-fit ODR line (solid color); the dashed gray line represents the 1:1 relationship. The bottom row displays the residuals ($\Delta$[Fe/H] = This Work - Literature). Robust statistics, including the median and the (scaled) MAD, are provided for each dataset. All panels use a shared vertical scale for direct comparison.}
    \label{fig:comparison_fits}
\end{figure}

The results of the ODR fits, illustrated in Fig.~\ref{fig:comparison_fits}, yield the following recalibration equations:

\begin{equation}
    \text{[Fe/H]}_{\text{T24b}} = (1.315 \pm 0.163) \times \text{[Fe/H]}_{\text{K22}} - (0.164 \pm 0.027)
\end{equation}

\begin{equation}
    \text{[Fe/H]}_{\text{T24b}} = (1.115 \pm 0.050) \times \text{[Fe/H]}_{\text{DS22}} - (0.049 \pm 0.012)
\end{equation}

The residuals for both fits are centered around zero, with dispersions  characterised by a (scaled) MAD of the order of 0.09 dex and 0.19 dex for \citet{kov22} and \citet{das22}, respectively.

\section{Coefficients of the photometric relation}\label{app:coefficients}
In Table~\ref{tab:feh_coeff}, we present the multivariate coefficients and their associated uncertainties for both our primary OLS relation and the alternative WLS calibration. As discussed in Sect.~\ref{sec-feh-fit}, the WLS empirical equation has the same functional form as Eq.~\ref{eq:dcepf-feh} but converges on a different subset of statistically significant predictors. Although the reported uncertainties for both models do not strictly follow conventional significant-digit rounding rules, we intentionally provide the coefficients with extended precision. This preserves the numerical accuracy of the higher-order polynomial terms and prevents round-off errors when the relations are implemented in computational applications.
\begin{table}
\caption{Best-fit coefficients and uncertainties for the photometric $\rm [Fe/H]$ relations. Columns $\beta^{\rm OLS}_k$ and $\sigma_{\beta^{\rm OLS}_k}$ report the parameters for our primary unweighted OLS relation (Eq.~\ref{eq:dcepf-feh}). Columns $\beta^{\rm WLS}_k$ and $\sigma_{\beta^{\rm WLS}_k}$ list the parameters obtained from the alternative WLS method. Dashes ($-$) indicate terms that were excluded from the respective according to the statistical significance.}
\label{tab:feh_coeff}
\centering
\begin{tabular}{lcccc}
\hline\hline
Term $p_k$ & $\beta^{OLS}_k$ & $\sigma_{\beta^{OLS}_k}$ & $\beta^{WLS}_k$ & $\sigma_{\beta^{WLS}_k}$ \\
\hline
intercept & 0.138 &  0.103 & -0.777  &   0.405\\
$\left(R_{31}^{G}\right)^{4}$ & 9.98 & 1.60 &-&-\\
$\left(R_{31}^{G}\right)^{2}$ & -7.355 & 0.881 &-7.522  &   1.031\\
$\left(A_{1}^{G}\right)^{4}$ & 8.59 & 1.27 & 85.55 &  33.88\\
$\left(A_{1}^{G}\right)^{2}$ & -8.37 & 1.14 & -45.96  & 13.10 \\
$\left(A_{1}^{G}\right)$ & - & - & 11.89   &  4.32 \\
$\left(R_{21}^{RP}\right)^{2}$ & 3.221 & 0.747 & 3.118  & 1.522\\
$\left(R_{21}^{RP}\right)$ & -4.959 & 0.908 &-5.382  &   1.457\\
$\left(A_{1}^{RP}\right)^{2}$ & 4.85 & 1.70 & 6.69 & 2.86\\
$\left(A_{2}^{RP}\right)^{2}$ & -112.5 & 13.0 & -85.3 & 34.5\\
$\left(A_{2}^{RP}\right)$ & 29.25 & 3.99 & 28.48   &  7.78\\
$\left(A_{3}^{RP}\right)^{4}$ & 1441 & 192 & - & -\\
\hline
\end{tabular}
\tablefoot{
Coefficients are reported with extended numerical precision to facilitate
their implementation in numerical codes. Uncertainties correspond to
formal 1$\sigma$ errors from the fit.
}
\end{table}

\section{Reference sample: Spectroscopic and Fourier Dataset}\label{app:Reference sample}
In this Appendix, we present the comprehensive dataset compiled for the Classical Cepheids calibration sample. Table ~\ref{tab:ref_sample_param} lists the Gaia DR3 source identifiers, pulsation periods (P), and spectroscopic [Fe/H] reference abundances, alongside the specific light curve quality metrics (number of data points nPt, phase uniformity indexes UI, rms scatter of the residuals in all bands and the standard deviation of the observations in all bands) and the selected Fourier parameters ($\rm R_{31}^G$, $\rm A_1G$, $\rm R_{21}^{RP}$, $\rm A_1{RP}$, $\rm A_2^{RP}$, $\rm A_3^{RP}$) utilized in our empirical metallicity calibration.
\begin{table*}
  \centering
  \caption{Spectroscopic and photometric parameters for the calibration sample of fundamental mode Classical Cepheids. The columns denote:
(1) Gaia DR3 source identifier (sourceid);
(2) pulsation period in days (P);
(3) to (5) number of photometric data points in the G, G$\rm _{BP}$, and G${_{RP}}$ bands (nPt$^G$, nPt$^{BP}$, nPt$^{RP}$), respectively;
(6) to (8) phase Uniformity Index for each band (UI$^G$, UI$^{BP}$, UI$^{RP}$) evaluating the phase coverage of the time series;
(9) reference high-resolution spectroscopic iron abundance ([Fe/H]);
(10) to (15) selected Fourier parameters entering the final empirical calibration relation ($\rm R_{31}^G$, $\rm A_1^G$, $\rm R_{21}^{RP}$, $\rm A_1^{RP}$, $\rm A_2^{RP}$, $\rm A_3^{RP}$);
(16) to (18) root-mean-square scatter of the residuals around the respective truncated Fourier fits (rms$^G$, rms$^{BP}$, rms$^{RP}$); (19) to (21) the standard deviation of the observations respectively in the G, BP and RP bands.}
  \label{tab:ref_sample_param} 
  \setlength{\tabcolsep}{3pt} 
\begin{tabular}{ccccccccccc}
  \hline
  sourceid & P & $\rm nPt^G$ & $\rm nPt^{BP}$ & $\rm nPt^{RP}$ & $\rm UI^G$ & $\rm UI^{BP}$ & $\rm UI^{RP}$ & [Fe/H] & $\rm R_{31}^G$ & $\rm A_1^G$ ...\\
    (1) & (2) & (3) & (4) & (5) &  (6)& (7) &(8)  &(9)  &(10)  &  (11) \\
  \hline
174489098011145216 & 6.46349 & 51 & 49 & 47 & 0.961 & 0.961 & 0.961 & 0.03 $\pm$ 0.19 & 0.142 $\pm$ 0.040 & 0.275 $\pm$ 0.012 ...\\
   200708636406382720 & 11.624 & 44 & 44 & 44 & 0.967 & 0.967 & 0.967 & -0.10 $\pm$ 0.19 & 0.114 $\pm$ 0.008 & 0.244 $\pm$ 0.002 ...\\
  261548119462093568 & 8.0023 & 47 & 47 & 47 & 0.951 & 0.951 & 0.951 & 0.08 $\pm$ 0.09 & 0.145 $\pm$ 0.010 & 0.249 $\pm$ 0.001 ... \\
  279382060625871360 & 3.29487 & 56 & 56 & 55 & 0.986 & 0.986 & 0.986 & -0.08 $\pm$ 0.19 & 0.134 $\pm$ 0.004 & 0.218 $\pm$ 0.001 ...\\
  377753613615880832 & 3.20092 & 43 & 46 & 45 & 0.968 & 0.976 & 0.974 & -0.29 $\pm$ 0.20 & 0.224 $\pm$ 0.015 & 0.261 $\pm$ 0.003 ... \\
  \multicolumn{11}{c}{...}\\
  \hline\\
\end{tabular}

\begin{tabular}{cccccccccc}
  \hline
  $\rm R_{21}^{RP}$ & $\rm A_1^{RP}$ & $\rm A_2^{RP}$ & $\rm A_3^{RP}$ & $\rm rms^G$ & $\rm rms^{BP}$ & $\rm rms^{RP}$ & $\rm \sigma^{G}$ &  $\rm \sigma^{BP}$ &  $\rm \sigma^{RP}$\\
      (12) & (13) & (14) & (15) & (16) &  (17)& (18) & (19) & (20) & (21)\\
  \hline
 ...0.377 $\pm$ 0.009 & 0.232 $\pm$ 0.002 & 0.087 $\pm$ 0.003 & 0.039 $\pm$ 0.002 & 0.002 & 0.003 & 0.003 & 0.199 & 0.162 & 0.136 \\
 ...0.123 $\pm$ 0.010 & 0.192 $\pm$ 0.002 & 0.024 $\pm$ 0.002 & 0.022 $\pm$ 0.002 & 0.002 & 0.003 & 0.004  & 0.210 & 0.158 & 0.137 \\
 ...0.227 $\pm$ 0.010 & 0.203 $\pm$ 0.002 & 0.046 $\pm$ 0.002 & 0.028 $\pm$ 0.003 & 0.002 & 0.003 & 0.002   & 0.227 & 0.185 & 0.149 \\
...0.322 $\pm$ 0.004 & 0.172 $\pm$ 0.001 & 0.055 $\pm$ 0.001 & 0.025 $\pm$ 0.001 & 0.003 & 0.006 & 0.003  & 0.203 & 0.164 & 0.128\\
 ...0.426 $\pm$ 0.009 & 0.198 $\pm$ 0.002 & 0.084 $\pm$ 0.002 & 0.043 $\pm$ 0.002 & 0.005 & 0.008 & 0.004  & 0.261 & 0.220 & 0.161\\
  \multicolumn{7}{c}{...}\\
  \hline
\end{tabular}
  \end{table*}


\begin{thebibliography}{}
\bibitem[Bhardwaj et al.(2017)]{bha17} Bhardwaj, A., Kanbur, S.~M., Marconi, M., et al.\ 2017, \mnras, 466, 3, 2805. doi:10.1093/mnras/stw3256
\bibitem[Bhardwaj et al.(2024)]{bha24} Bhardwaj, A., Ripepi, V., Testa, V., et al.\ 2024, \aap, 683, A234. doi:10.1051/0004-6361/202348140
\bibitem[Boggs \& Rogers(1990)]{Boggs1990} Boggs, P.~T., \& Rogers, J.~E.\ 1990, Contemporary Mathematics, 112, 186
\bibitem[Breiman(2001)]{Breiman2001} Breiman, L.\ 2001, Machine Learning, 45, 5. doi:10.1023/A:1010933404324
\bibitem[Breuval et al.(2022)]{bre22} Breuval, L., Riess, A.~G., Kervella, P., et al.\ 2022, \apj, 939, 2, 89. doi:10.3847/1538-4357/ac97e2
\bibitem[Breuval et al.(2024)]{bre24} Breuval, L., Riess, A.~G., Casertano, S., et al.\ 2024, \apj, 973, 1, 30. doi:10.3847/1538-4357/ad630e
\bibitem[Breuval et al.(2025)]{bre25} Breuval, L., Anand, G.~S., Anderson, R.~I., et al.\ 2025, \apj, 994, 1, 111. doi:10.3847/1538-4357/ae0cb9
\bibitem[Catanzaro et al.(2024)]{cat24} Catanzaro, G., Ripepi, V., Salaris, M., et al.\ 2024, \aap, 682, L21. doi:10.1051/0004-6361/202449160
\bibitem[Clementini et al.(2023)]{cle23} Clementini, G., Ripepi, V., Garofalo, A., et al.\ 2023, \aap, 674, A18. doi:10.1051/0004-6361/202243964
\bibitem[Danielski et al.(2018)]{dan18} Danielski, C., Babusiaux, C., Ruiz-Dern, L., et al.\ 2018, \aap, 614, A19. doi:10.1051/0004-6361/201732327
\bibitem[da Silva et al.(2022)]{das22} da Silva, R., Crestani, J., Bono, G., et al.\ 2022, \aap, 661, A104. doi:10.1051/0004-6361/202142957
\bibitem[De Somma et al.(2022)]{des22} De Somma, G., Marconi, M., Molinaro, R., et al.\ 2022, \apjs, 262, 1, 25. doi:10.3847/1538-4365/ac7f3b
\bibitem[D{\'e}k{\'a}ny et al.(2021)]{dek21} D{\'e}k{\'a}ny, I., Grebel, E.~K., \& Pojma{\'n}ski, G.\ 2021, \apj, 920, 1, 33. doi:10.3847/1538-4357/ac106f
\bibitem[Gaia Collaboration et al.(2016)]{gaia2016} Gaia Collaboration, Prusti, T., de Bruijne, J.~H.~J., et al.\ 2016, \aap, 595, A1. doi:10.1051/0004-6361/201629272
\bibitem[Genovali et al.(2014)]{gen14} Genovali, K., Lemasle, B., Bono, G., et al.\ 2014, \aap, 566, A37. doi:10.1051/0004-6361/201323198
\bibitem[Genovali et al.(2015)]{gen15} Genovali, K., Lemasle, B., da Silva, R., et al.\ 2015, \aap, 580, A17. doi:10.1051/0004-6361/201525894
\bibitem[Grisoni et al.(2018)]{gri18} Grisoni, V., Spitoni, E., \& Matteucci, F.\ 2018, \mnras, 481, 2, 2570. doi:10.1093/mnras/sty2444
\bibitem[Hocd{\'e} et al.(2023)]{hoc23} Hocd{\'e}, V., Smolec, R., Moskalik, P., et al.\ 2023, \aap, 671, A157. doi:10.1051/0004-6361/202245038
\bibitem[Iorio \& Belokurov(2021)]{ior21} Iorio, G. \& Belokurov, V.\ 2021, \mnras, 502, 4, 5686. doi:10.1093/mnras/stab005
\bibitem[Jurcsik \& Kovacs(1996)]{jur96} Jurcsik, J. \& Kovacs, G.\ 1996, \aap, 312, 111. 
\bibitem[Kovtyukh et al.(2022)]{kov22} Kovtyukh, V., Lemasle, B., Bono, G., et al.\ 2022, \mnras, 510, 2, 1894. doi:10.1093/mnras/stab3530
\bibitem[Klagyivik et al.(2013)]{kla13} Klagyivik, P., Szabados, L., Szing, A., et al.\ 2013, \mnras, 434, 3, 2418. doi:10.1093/mnras/stt1176
\bibitem[Kovacs \& Zsoldos(1995)]{kov95} Kovacs, G. \& Zsoldos, E.\ 1995, \aap, 293, L57. 
\bibitem[Leavitt \& Pickering(1912)]{lea12} Leavitt, H.~S. \& Pickering, E.~C.\ 1912, Harvard College Observatory Circular, 173, 1. 
\bibitem[Lemasle et al.(2013)]{lem13} Lemasle, B., Fran{\c{c}}ois, P., Genovali, K., et al.\ 2013, \aap, 558, A31. doi:10.1051/0004-6361/201322115
\bibitem[Liaw A, Wiener M (2002)]{lia02} Liaw A, Wiener M (2002). Classification and Regression by randomForest. R News, *2*(3), 18-22. <https://CRAN.R-project.org/doc/Rnews/>.
\bibitem[Long et al.(2025)]{lon25} Long, G., Yuan, H., Xu, S., et al.\ 2025, \apj, 983, 1, 51. doi:10.3847/1538-4357/adbf93
\bibitem[Marconi et al.(2024)]{mar24} Marconi, M., De Somma, G., Molinaro, R., et al.\ 2024, \mnras, 529, 4, 4210. doi:10.1093/mnras/stae734
\bibitem[Mullen et al.(2021)]{mul21} Mullen, J.~P., Marengo, M., Mart{\'\i}nez-V{\'a}zquez, C.~E., et al.\ 2021, \apj, 912, 2, 144. doi:10.3847/1538-4357/abefd4
\bibitem[Muraveva et al.(2025)]{mur25} Muraveva, T., Giannetti, A., Clementini, G., et al.\ 2025, \mnras, 536, 3, 2749. doi:10.1093/mnras/stae2679
\bibitem[Planck Collaboration et al.(2020)]{planck2020} Planck Collaboration, Aghanim, N., Akrami, Y., et al.\ 2020, \aap, 641, A6. doi:10.1051/0004-6361/201833910
\bibitem[Pont et al.(2001)]{pon01} Pont, F., Kienzle, F., Gieren, W., et al.\ 2001, \aap, 376, 892. doi:10.1051/0004-6361:20010377
\bibitem[Riess et al.(2022)]{rie22} Riess, A.~G., Yuan, W., Macri, L.~M., et al.\ 2022, \apjl, 934, 1, L7. doi:10.3847/2041-8213/ac5c5b
\bibitem[Riess et al.(2024)]{rie24} Riess, A.~G., Scolnic, D., Anand, G.~S., et al.\ 2024, \apj, 977, 1, 120. doi:10.3847/1538-4357/ad8c21
\bibitem[Ripepi et al.(2026)]{rip26} Ripepi, V., Trentin, E., Catanzaro, G., et al.\ 2026, \aap, 708, A216. doi:10.1051/0004-6361/202556963
\bibitem[Ripepi et al.(2025)]{rip25} Ripepi, V., Trentin, E., Catanzaro, G., et al.\ 2025, , arXiv:2508.17447. doi:10.48550/arXiv.2508.17447
\bibitem[Ripepi et al.(2023)]{rip23} Ripepi, V., Clementini, G., Molinaro, R., et al.\ 2023, \aap, 674, A17. doi:10.1051/0004-6361/202243990
\bibitem[Ripepi et al.(2022)]{rip22} Ripepi, V., Catanzaro, G., Molinaro, R., et al.\ 2022, \mnras, 516, 2, 2887. doi:10.1093/mnras/stac1787
\bibitem[Ripepi et al.(2021)]{rip21} Ripepi, V., Catanzaro, G., Molinaro, R., et al.\ 2021, \mnras, 508, 3, 4047. doi:10.1093/mnras/stab2460
\bibitem[Romaniello et al.(2008)]{rom08} Romaniello, M., Primas, F., Mottini, M., et al.\ 2008, \aap, 488, 2, 731. doi:10.1051/0004-6361:20065661
\bibitem[Romaniello et al.(2022)]{rom22} Romaniello, M., Riess, A., Mancino, S., et al.\ 2022, \aap, 658, A29. doi:10.1051/0004-6361/202142441
\bibitem[Satopaa(2011)]{sat11} Satopaa, V.\ 2011, DOI:10.1109/ICDCSW.2011.20 Distributed Computing Systems Workshops (ICDCSW), 2011 31st International 
\bibitem[Schmidt(1991)]{sch91} Schmidt, E.~G.\ 1991, \aj, 102, 1766. doi:10.1086/115999
\bibitem[Scowcroft et al.(2016)]{sko16} Scowcroft, V., Seibert, M., Freedman, W.~L., et al.\ 2016, \mnras, 459, 2, 1170. doi:10.1093/mnras/stw628
\bibitem[Simon \& Lee(1981)]{sim81} Simon, N.~R. \& Lee, A.~S.\ 1981, \apj, 248, 291. doi:10.1086/159153
\bibitem[Simon(1988)]{sim88} Simon, N.~R.\ 1988, \apj, 328, 747. doi:10.1086/166333
\bibitem[Szabados \& Klagyivik(2012)]{sza12} Szabados, L. \& Klagyivik, P.\ 2012, \aap, 537, A81. doi:10.1051/0004-6361/201117815
\bibitem[Szil{\`a}di et al.(2018)]{szi18} Szil{\`a}di, K., Vink{\`o}, J., \& Szabados, L.\ 2018, \actaa, 68, 2, 111. doi:10.32023/0001-5237/68.2.2
\bibitem[Trentin et al.(2026)]{tre26} Trentin, E., Catanzaro, G., Ripepi, V., et al.\ 2026, \aap, 707, A142. doi:10.1051/0004-6361/202557996
\bibitem[Trentin et al.(2024a)]{tre24a} Trentin, E., Catanzaro, G., Ripepi, V., et al.\ 2024, \aap, 690, A246. doi:10.1051/0004-6361/202450376
\bibitem[Trentin et al.(2024b)]{tre24b} Trentin, E., Ripepi, V., Molinaro, R., et al.\ 2024, \aap, 681, A65. doi:10.1051/0004-6361/202347195
\bibitem[Trentin et al.(2023)]{tre23} Trentin, E., Ripepi, V., Catanzaro, G., et al.\ 2023, \mnras, 519, 2, 2331. doi:10.1093/mnras/stac2459
\bibitem[Zsoldos(1995)]{zso95} Zsoldos, E.\ 1995, IAU Colloquium 155: Astrophysical Applications of Stellar Pulsation, 83, 351. 


\end{thebibliography}
\end{document}